Routledge
Taylor & Francis Group

Check for updates

# Strengthening legal protection against discrimination by algorithms and artificial intelligence

Frederik J. Zuiderveen Borgesius  [a,b]

[a]Institute for Computing and Information Sciences (iCIS), Radboud University, Nijmegen, Netherlands; [b]Interdisciplinary Hub for Security, Privacy and Data Governance (iHub), Radboud University, Nijmegen, Netherlands

**ABSTRACT**

Algorithmic decision-making and other types of artificial intelligence (AI) can be used to predict who will commit crime, who will be a good employee, who will default on a loan, etc. However, algorithmic decision-making can also threaten human rights, such as the right to non-discrimination. The paper evaluates current legal protection in Europe against discriminatory algorithmic decisions. The paper shows that non-discrimination law, in particular through the concept of indirect discrimination, prohibits many types of algorithmic discrimination. Data protection law could also help to defend people against discrimination. Proper enforcement of non-discrimination law and data protection law could help to protect people. However, the paper shows that both legal instruments have severe weaknesses when applied to artificial intelligence. The paper suggests how enforcement of current rules can be improved. The paper also explores whether additional rules are needed. The paper argues for sector-specific – rather than general – rules, and outlines an approach to regulate algorithmic decision-making.

## 1. Introduction

The use of algorithmic decision-making has become common practice across a wide range of sectors. We use algorithmic systems for spam filtering, traffic planning, logistics management, diagnosing diseases, speech recognition, and much more. Although algorithmic decision-making can seem rational, neutral, and unbiased, it can also lead to unfair and illegal discrimination.

The two main questions for this paper are as follows. (i) Which legal protection against algorithmic discrimination exists in Europe, and what are its limitations? (ii) How could that legal protection be improved? The first research question for this paper is evaluative, as it evaluates current law. The second research question can be characterised as a design question, as it discusses whether, and if so how, new laws should be designed.[1]

The paper focuses on the two most relevant legal instruments for defending people against algorithmic discrimination: non-discrimination law and data protection law. The paper speaks of 'discrimination' when referring to objectionable or illegal discrimination, for example on the basis of gender, sexual preference, or ethnic origin. The word 'differentiation' refers to discrimination, or making distinctions, in a neutral, unobjectionable, sense.[2]

The paper's main contributions to scholarship are the following. First, there has not been much legal analysis of European non-discrimination law in the context of algorithmic decision-making. The few papers that discuss European non-discrimination law do so with a focus on EU law;[3] this paper discusses the norms from the European Convention on Human Rights. Second, building on other literature, the paper assesses how data protection law can help to protect people against discrimination. Third, the paper proposes an approach to regulate algorithmic decision-making in a sector-specific way. The paper could be useful for scholars, practitioners, and for policymakers that want to regulate algorithmic decision-making.

The paper focuses on the overarching rules in Europe (the region of the Council of Europe, with 47 member states); national rules are out of scope.[4] Because of the focus on discrimination, questions relating to, for instance, privacy and freedom of expression are outside the scope of the paper. The paper is based on, and includes text from, a report by the author for the Anti-discrimination department of the Council of Europe.[5]

The paper is structured as follows. Section 2 introduces algorithmic decision-making, artificial intelligence, and some related concepts. Section 3 shows that there is a problem, and gives examples of algorithmic decision-making that leads, or could lead, to discrimination. Section 4 turns to law. The paper discusses current legal protection against algorithmic discrimination, and flags strengths and weaknesses of that protection. Section 5 suggests how enforcement of current non-discrimination norms can be improved. The section also explores whether algorithmic decision-making necessitates amending non-discrimination norms. The paper outlines an approach to adopting rules regarding algorithmic decision-making. Section 6 offers concluding thoughts.

## 2. Algorithmic decision-making and artificial intelligence

This section introduces algorithmic decision-making, AI and some related concepts. While there is no consensus about defining algorithmic decision-making and AI, some descriptions may be helpful. An algorithm is 'an abstract, formalized description of a computational procedure'.[6] In this paper, 'decision' refers to the output of that procedure. Roughly speaking, an algorithm can be seen as a computer program. Algorithmic decision-making thus refers to the process by which an algorithm produces an output.

Sometimes, an algorithm decides in a fully automated fashion. For example, a spam filter can filter out, automatically, spam messages from one's email account. Sometimes, decisions are *partly* automated: humans make decisions assisted by algorithms. For example, a bank employee may decide whether a customer can borrow money from the bank, after an algorithmic system assessed the customer's creditworthiness.

When discussing discrimination however, many risks are similar for fully and partly automated decisions. People might follow recommendations by computers, because those recommendations seem rational or infallible,[7] or because people try to minimise their own responsibility.[8] (Below we see that some legal rules do distinguish fully and partly automated decisions.[9])

Artificial intelligence (AI) can be described, in the words of a famous computer science textbook, as 'the study of the design of intelligent agents'.[10] An agent is 'something that acts', such as a computer.[11] One type of AI has been particularly successful the past decade: machine learning.[12] As Lerh and Ohm explain, 'machine learning refers to an automated process of discovering correlations (sometimes alternatively referred to as relationships or patterns) between variables in a dataset, often to make predictions or estimates of some outcome'.[13] Largely because of the availability of enormous amounts of data, machine learning has become widely used during the past decade. Machine learning is so successful, that nowadays many people use the phrases AI and machine learning interchangeably.[14]

This paper sacrifices precision for readability, and uses, 'algorithmic decision-making', 'AI', etc., without specifying whether the phrases refer to machine learning or another technology. Hence, in this paper, an 'algorithmic system' can refer to, for example, a computer running an algorithm that was fed data by its human operators.

## 3. Discrimination risks of algorithmic decision-making

This section shows that there is a problem, and gives examples of algorithmic decision-making with discriminatory effects. Algorithmic decisions can be discriminatory, for instance when the system learnt from discriminatory human decisions.[15]

A notorious example of an algorithmic system with discriminatory effects is COMPAS, which stands for 'Correctional Offender Management Profiling for Alternative Sanctions'.[16] COMPAS is used in parts of the US to predict whether defendants will commit crime again. COMPAS is meant to help judges to determine whether somebody should be allowed to go on probation (supervision outside prison). COMPAS does not use ethnic origin or skin colour as an input. Yet, research by Angwin et al. showed in 2016 that COMPAS is 'biased against blacks'.[17]

COMPAS 'correctly predicts recidivism 61 percent of the time', note Angwin et al. However,

blacks are almost twice as likely as whites to be labeled a higher risk but not actually re-offend. It makes the opposite mistake among whites: They are much more likely than blacks to be labeled lower risk but go on to commit other crimes.[18]

Moreover,

Black defendants were also twice as likely as white defendants to be misclassified as being a higher risk of violent recidivism. And white violent recidivists were 63 percent more likely to have been misclassified as a low risk of violent recidivism, compared with black violent recidivists.[19]

Algorithmic systems can also be used for predictive policing, 'the application of analytical techniques – particularly quantitative techniques – to identify likely targets for police intervention and prevent crime or solve past crimes by making statistical predictions'.[20]

However, predictive policing can reproduce and even amplify existing discrimination. Say the police pay extra attention in a neighbourhood with many immigrants, while that neighbourhood has average crime levels. The police register more crime in that neighbourhood than elsewhere. Because the numbers show more crime is registered (and thus seems to occur) in that the neighbourhood, even more police are sent there. This way, policing on the basis of crime statistics can cause a feedback loop.[21]

In the private sector too, algorithmic decision-making can have discriminatory effects. For instance, AI can be used by firms to select employees; or by schools to select students. Biased training data, however, could lead to discriminatory decisions. The training data can be biased because they represent discriminatory human decisions. The risk can be illustrated with an example from the 1980s in the UK.[22] A medical school used an algorithmic system to select students from the many student applications. The system was trained on the admission files from earlier years, when the school's employees selected which applicants could enter medical school. The training data showed the system which characteristics (the input) correlated with the desired output (being admitted to the medical school). The algorithmic system reproduced that selection method.

The computer program discriminated against women and people with an immigrant background. In the years that provided the training data, the people that selected the students were biased against women and people with an immigrant background. As the British medical journal noted, 'the program was not introducing new bias but merely reflecting that already in the system'.[23] In sum, if an algorithmic system is trained on biased data, the system risks reproducing that bias.

Targeted online advertising is largely driven by algorithmic decision-making. Such advertising is a profitable sector for some companies. For example, Facebook and Google make most of their money from online advertising. However, online advertising can have discriminatory effects.

Datta et al. simulated internet users and made them self-declare as male or female. The researchers analysed the ads presented by Google to the simulated internet users.[24] 'Google showed the simulated males ads from a certain career coaching agency that promised large salaries more frequently than the simulated females, a finding suggestive of discrimination.'[25] Because of the lack of transparency of online advertising, it is unclear why women were shown fewer ads for high-paying jobs.[26] Their research illustrates that the opaqueness of algorithmic decision-making can make it harder to discover discrimination and its cause.

Image search systems can also have discriminatory effects. In 2016, Google Images showed many mugshots when people searched for 'three black teenagers'. In contrast, Google Images showed pictures of happy white kids when people searched for 'three white teenagers'. Google said: 'Our image search results are a reflection of content from across the web, including the frequency with which types of images appear and the way they're described online'.[27] It could indeed be argued that Google's algorithmic system merely reflected society.[28] However, even if the fault lies with society rather than with the algorithmic system, those image search results could influence people's beliefs.

Research by Kay, Matuszek and Munson showed that 'image search results for occupations slightly exaggerate gender stereotypes and portray the minority gender for an occupational less professionally. There is also a slight underrepresentation of women'.[29] In sum, in the public and the private sector, algorithmic decision-making can lead to discriminatory effects.

While decisions by algorithms can have discriminatory effects, algorithms are not inherently bad or discriminatory. Algorithms might still perform better than human decision-makers. Many humans also discriminate. Indeed, sometimes algorithmic systems discriminate because they reproduce discrimination by humans.[30] Hence, it makes a difference whether one compares algorithmic decision-making with human decisions in the real world (which are sometimes discriminatory), or with hypothetical decisions in an ideal world without discrimination. Of course, we should aim for a world without any unfair or illegal discrimination.

Algorithmic decision-making can also be used to fight discrimination.[31] Suppose an algorithmic system shows that a collection of stock photos contains gender stereotypes. One interpretation is that the algorithmic system illustrates stereotyped behaviour that already exists. Hence, an algorithmic system could help to discover existing discrimination that would otherwise have remained hidden.

There is a vibrant new subfield in computer science, focusing on fairness, accountability, and transparency in the context of algorithmic decision-making.[32] Computer scientists investigate, for instance, ways to discover and prevent discrimination, and ways to build non-discrimination norms into algorithmic systems.

## 4. Current regulation

### 4.1. Non-discrimination law and the protection it offers against algorithmic discrimination

Non-discrimination law and data protection law are the main legal instruments that could protect the right to non-discrimination in the context of algorithmic decision-making. Both instruments are introduced in this section. The paper focuses on the main principles of the law, omitting greater detail.[33] We start with the right to non-discrimination.

Many constitutions and human rights treaties prohibit discrimination.[34] For example, the European Convention on Human Rights states in article 14:

> The enjoyment of the rights and freedoms set forth in this Convention shall be secured without discrimination on any ground such as sex, race, colour, language, religion, political or other opinion, national or social origin, association with a national minority, property, birth or other status.[35]

Case law shows that the European Convention on Human Rights prohibits both *direct* and *indirect* discrimination.[36] In brief, direct discrimination means discriminating on the basis of a protected ground, such as ethnic origin. Direct discrimination is described as follows by the European Court of Human Rights: 'there must be a difference in the treatment of persons in analogous, or relevantly similar, situations', which is based 'on an identifiable characteristic'.[37] A similar definition is used in EU non-discrimination law.[38]

Indirect discrimination means that a practice which seems neutral at first glance ends up discriminating against people of a certain ethnic origin, or another protected characteristic.[39] (Indirect discrimination is called 'disparate impact' in the United States; direct discrimination is called disparate treatment.)

The European Court of Human Rights describes indirect discrimination as follows:

> a difference in treatment may take the form of disproportionately prejudicial effects of a general policy or measure which, though couched in neutral terms, discriminates against a group. Such a situation may amount to 'indirect discrimination', which does not necessarily require a discriminatory intent.[40]

EU law uses a similar definition of indirect discrimination.[41]

For indirect discrimination, it is not relevant whether the alleged discriminator intended to discriminate. It is the effect of a practice that counts.[42] Therefore, discriminators cannot evade the prohibition of indirect discrimination by proving that they did not mean to discriminate.

We saw that algorithmic systems can cause discrimination. For example, algorithmic decision-making could breach the prohibition of indirect discrimination if an algorithmic system rejects job application letters from a disproportionate number of people with a certain ethnicity. Additionally, organisations could use algorithms to discriminate on purpose. For instance, a firm that wanted to discriminate against people of a certain ethnicity could discriminate on the basis of a proxy that correlates with that ethnicity, such as postal code: a type of algorithmic red-lining.

Non-discrimination law could help to protect people against algorithmic discrimination. Still, non-discrimination law has several weaknesses when applied to algorithmic discrimination. For instance, a victim must show that a seemingly neutral rule, practice, or decision disproportionately affects a protected class and is thus *prima facie* discriminatory.[43]

However, indirect discrimination can remain hidden to both the organisation and the victim. Suppose that somebody applies for a loan on the website of a bank. The bank uses an algorithmic system to decide on such requests. If the bank automatically denies a loan to a customer on its website, the customer does not see why the loan was denied. And the customer cannot see whether the bank's algorithmic system denies loans to a disproportionate percentage of, for example, women.[44] Hence, even if customers knew that an algorithm rather than a bank employee decided, it would be difficult for them to discover whether the algorithm is discriminatory.

Algorithmic systems are often 'black boxes', for several reasons.[45] For instance, most people lack the technical expertise to understand how such systems arrive at decisions. Even the experts who have built an algorithmic system may not know how that system will behave when used in practice and is fed certain data. In addition, trade or state secrets may hinder obtaining information about algorithmic systems.[46] Because of the lack of transparency of algorithmic decisions, it is hard for people to assess whether they were discriminated against.

Another problem with non-discrimination law is that the prohibition of indirect discrimination does not provide a clear black-and-white rule. The concept of indirect discrimination results in rather open-ended standards, which are often difficult to apply in practice. The prohibition of indirect discrimination does not apply if the alleged discriminator successfully invokes an objective justification. The European Court of Human Rights says:

> A general policy or measure that has disproportionately prejudicial effects on a particular group may be considered discriminatory even where it is not specifically aimed at that group and there is no discriminatory intent. This is only the case, however, if such policy or measure has no 'objective and reasonable' justification.[47]

However, the justification must be objective and reasonable. In addition, the practice (measure, or policy) must be proportional to that goal. The European Court of Human Rights states that a practice does not meet those requirements if that practice 'has no objective and reasonable justification, that is if it does not pursue a legitimate aim or if there is not a reasonable relationship of proportionality between the means employed and the aim sought to be achieved'.[48]

It depends on all the circumstances of a case whether an alleged discriminator can invoke such an objective justification.[49] Therefore, it is not always clear whether a certain practice breaches the prohibition of indirect discrimination.

In sum, non-discrimination law prohibits many discriminatory effects of algorithmic decision-making, in particular through the concept of indirect discrimination. Enforcement is difficult, however, and non-discrimination law has weaknesses. Could data protection law help? We turn to that question now.

## 4.2. Data protection law and the protection it offers against algorithmic discrimination

The European Convention on Human Rights (1950) grants people a right to private life.[50] The European Court of Human Rights has deduced data protection principles from the Convention's right to private life. The court thus gives certain data protection principles a human rights-like status.[51]

In the European Union, the right to protection of personal data has the status of a human right. All European Union member states are members of the Council of Europe and subject to the European Convention on Human Rights (1953). But the EU also has its own Charter of Fundamental Rights of the European Union (2000, legally binding since 2009). The Charter grants people, in addition to a right to private life, a separate right to the protection of personal data.[52] (For the purposes of this paper, the phrases 'fundamental rights' and 'human rights' can be regarded as synonyms.)[53]

More detailed rules can be found in the Council of Europe's Data Protection Convention 108+ (revised in 2018[54]) and the European Union's General Data Protection Regulation (GDPR, from 2016). The main rules in both instruments are similar. For brevity, we speak of 'data protection law' below, to refer to those main rules.

Data protection law aims to defend fairness and human rights when organisations use personal data.[55] Data protection law grants rights to people whose data are being used (data subjects),[56] and imposes obligations on organisations that use personal data (data controllers).[57]

The core of data protection law consists of eight principles, which can be summarised as follows: (a) Personal data may only be processed lawfully, fairly and transparently. (b) Such data may only be collected for a purpose that is specified in advance, and should only be used for purposes that are compatible with the original purpose. (c) Organisations should not collect or use more data than necessary. (d) Organisations must ensure that such data are sufficiently accurate and up to date. (e) Organisations should not store the data for an unreasonably long time. (f) Organisations must ensure data security.[58] (g) The organisation that determines the purposes and means for processing (the 'controller') is responsible for compliance.[59]

These principles lie at the foundation of the Data Protection Convention 108+ and the GDPR.[60] Over a hundred countries in the world have statutes with similar principles.[61] Below, we focus more on the GDPR than on Convention 108, because the GDPR provides more details.

Data protection law could be used to allay information asymmetry and to mitigate risks of unfair and illegal discrimination.[62] For example, data protection law requires that organisations are open and transparent about their use of personal data. Therefore, organisations must provide information, for instance in a privacy notice, about all stages of algorithmic decision-making that involve personal data.[63] Even though most people ignore privacy notices,[64] such notices could be helpful for researchers, journalists, and supervisory authorities, to obtain information about an organisation's practices.

In certain situations, organisations are required to conduct a data protection impact assessment (DPIA) under the GDPR and the Data Protection Convention 108. A DPIA can be described as a process to 'systematically analyse, identify and minimise the data protection risks of a project or plan'.[65] Scholars such as Goodman, Hacker, and Kaminski and Malgieri highlight the potential of DPIAs to protect people against algorithmic discrimination.[66]

Under the GDPR, an organisation must conduct a DPIA when a practice is 'likely to result in a high risk to the rights and freedoms of natural persons', especially when using new technologies.[67] In some circumstances, the GDPR assumes a high risk, for example when organisations take fully automated decisions that seriously affect people.[68] For many algorithmic systems that make decisions about people, the GDPR thus requires a DPIA.[69] When conducting a DPIA, organisations must also consider the risk of unfair or illegal discrimination.[70]

The Council of Europe and the EU require each member state to have an independent Data Protection Authority.[71] Such Data Protection Authorities must have powers of investigation.[72] Under the GDPR, a Data Protection Authority can, for example, access to premises of an organisation using personal data, and order an organisation to give access to its computers.[73]

### 4.3. Data protection law's rules on automated decisions and the protection they offer against algorithmic discrimination

The GDPR and Convention 108 contain specific rules for certain types of 'automated individual decision-making'.[74] These rules have several goals, including protecting people against unfair or illegal discrimination.[75] First, we take a look at the GDPR.

Article 22 of the GDPR is sometimes called the Kafka provision. In principle, Article 22 prohibits certain fully automated decisions with legal or similar significant effects. Article 22 applies, for example, to fully automated credit scoring and e-recruiting.[76] The main rule of article 22 reads as follows:

> The data subject shall have the right not to be subject to a decision based solely on automated processing, including profiling, which produces legal effects concerning him or her or similarly significantly affects him or her.[77]

That rule could be summarised as: organisations may not subject people to certain types of fully automated decisions with far-reaching effects. Article 22 states that people have a 'right not to be subject to' certain decisions. That right can be interpreted as a prohibition of such decisions.[78]

The main rule of article 22 can be broken down into four conditions: (i) there is a decision, that is based (ii) solely (iii) on automated data processing; (iv) the decision has legal or similarly significant effects for the data subject.[79]

What is a decision with 'legal effects'? An example is a court decision, or a decision regarding a social benefit granted by law, such as pension payments.[80] An example of a decision with 'similarly significant' effects is a decision by a bank to deny credit to somebody.[81]

The prohibition of certain automated decisions is not absolute; there are exceptions. In brief, the prohibition does not apply (i) if the individual gave consent to the automated decision; or if the decision is (ii) necessary for a contract between the individual and the data controller, or (iii) is authorised by law.[82]

A different rule is triggered if a controller can rely on the (i) consent or (ii) contract exception. That rule says:

> the data controller shall implement suitable measures to safeguard the data subject's rights and freedoms and legitimate interests, at least the right to obtain human intervention on the part of the controller, to express his or her point of view and to contest the decision.[83]

Thus, in some circumstances, people can demand the organisation to make an employee reconsider the automated decision. For example, a bank could ensure that customers can call the bank to have a human reconsider the decision, if the bank automatically denies them a loan through the bank's website.

The GDPR contains specific transparency requirements for automated decisions:

> [T]he controller shall provide the data subject with the following information (…) the existence of automated decision-making, including profiling (…) and, at least in those cases, meaningful information about the logic involved, as well as the significance and the envisaged consequences of such processing for the data subject.[84]

In some circumstances, an organisation must thus explain that it uses automated decision-making, and must provide 'meaningful information' about the underlying logic of that decision-making process.

There is a lively scholarly discussion about the extent to which the GDPR creates a 'right to explanation' of individual decisions. One the one hand, scholars such as Edwards and Veale, and Wachter et al. doubt the effectiveness of such a right, and point out that many algorithmic decisions remain outside the scope of the GDPR's rules.[85] For instance, it is debatable to what extent the GDPR's article 22 applies to decisions that are largely, rather than 'solely', based on automated processing. It could be argued that article 22 does not apply if a bank employee denies a loan on the basis of a recommendation by an algorithmic system.[86] On the other hand, scholars such as Malgieri and Comandé say that the GDPR does offer a right to explanation (or 'legibility') of automated decisions.[87]

It is too early to tell what the effect is of the GDPR rules on automated decisions. The predecessor of the GDPR provision on automated decisions has hardly been applied in practice. Still, even if the new provision on automated decisions will never be applied by judges, the provision had some effect already: it triggered an interdisciplinary discussion on explaining algorithmic decisions.

As Veale and Edwards note, the rules on automated decisions in Convention 108 are more generous for individuals. Under Convention 108, people have a right 'to obtain, on request, knowledge of the reasoning underlying data processing where the results of such processing are applied to him or her'.[88] Hence, Convention 108 does not limit such a right to decisions with legal or significant effects. Again, we have to wait to see what the effect will be of this different phrasing.

## 4.4. Limitations of data protection law in the context of algorithmic discrimination

Data protection law is not a silver bullet to protect people against algorithmic discrimination. Several caveats are necessary. First, there is a compliance and enforcement deficit. Many organisations did not take compliance with data protection law seriously. It appears that compliance improved with the arrival of the GDPR (which applies since 2018), but it is too early to assess whether compliance improved. Another problem is that Data Protection Authorities are overburdened. Moreover, many Data Protection Authorities lack the power to impose serious sanctions. In the EU, the GDPR granted such authorities new powers, but it is too early to say much about the effect of those powers.

Second, as data protection law only applies to personal data, algorithmic decision-making processes are partly outside the scope of data protection law. Data protection law does not apply to predictive models, because they do not relate to identifiable persons. For instance, a predictive model that says '80% of the people living in postal code 10017 pay their bills late' does not refer to an individual. Therefore, the model is not a personal datum. (Data protection law does apply when such a predictive model is applied to an individual.[89])

Third, data protection law uses many open and abstract norms, rather than clear and concrete rules. Data protection law must use open norms, because its provisions apply in many different situations, in the private and the public sector. This regulatory approach, an omnibus approach, has many advantages. For example, the open norms do not have to be adapted each time when a new technology is developed. A disadvantage of data protection law's omnibus approach is that open norms can be difficult to apply in practice.

A fourth caveat concerns data protection law's strict rules on 'special categories' of data (sometimes called 'sensitive data'), such as data regarding ethnic origin or health status.[90] As noted by De Schutter and Ringelheim, that stricter regime creates challenges when assessing and mitigating discrimination. Many of the methods to tackle discrimination in algorithmic systems implicitly assume that organisations hold these sensitive data – yet to meet data protection law, many organisations may not be. Tension remains between respecting data protection law and collecting sensitive data to fight discrimination.[91]

Fifth, even when assuming that people have a right to explanation regarding algorithmic decisions, it is often difficult, if not impossible, to explain the logic behind a decision, when an algorithmic system arrives at that decision after analysing large amounts of data.[92] Moreover, in some circumstances, an explanation might not be of much help to people, note Edwards and Veale.[93]

Nevertheless, it would be good if organisations offered more openness and explanations about their algorithmic decisions. Scholars such as Hildebrandt and Gutwirth have called for more openness regarding automated decision-making for at least a decade.[94] Computer scientists are exploring various ways of improving the transparency and explainability of algorithmic decisions.[95]

It is too early to assess the effects of Convention 108 and the GDPR. More legal research is needed on how data protection law could help mitigate discrimination risks.[96] While the right to the protection of personal data and data protection law are largely untested as non-discrimination tools, they offer possibilities to fight illegal discrimination.

### 4.5. Other regulation

Other legal instruments could also help to protect people against algorithmic discrimination. For instance, consumer law could help to defend people against manipulative algorithmic advertising.[97] Competition law could also be a useful. Discriminatory behaviour by a firm causes more problems when the firm holds a monopoly position.[98] In the public sector, criminal law and administrative law could help to safeguard the fairness of procedures.[99] Freedom of information laws could help journalists, researchers, and others to obtain information about algorithmic systems in the public sector.[100] However, the application of these fields of law to protect people in the area of algorithmic decision-making is largely unexplored. A discussion of those fields of law falls outside the scope of this paper.

Several organisations published principles for fair, accountable, or ethical AI. To illustrate: the Organisation for Economic Co-operation and Development (OECD) published a recommendation on AI,[101] and the Council of Europe published a draft recommendation on the human rights impacts of algorithmic systems.[102] The European Commission set up a High-Level Expert Group on Artificial Intelligence, which proposed AI Ethics Guidelines in 2019: a type of co-regulation.[103]

Other organisations published self-regulatory principles on ethics and AI. Examples include the (US-based) Future of Life Institute,[104] the UNI Global Union,[105] and Google.[106]

On the one hand, self-regulation is commendable. It can hardly be denied that ethical AI is preferable over unethical AI. Self-regulation could help mitigate discrimination, and could provide inspiration for legislators. On the other hand, there are serious problems with self-regulation. Most importantly: self-regulation is non-binding. Human rights protection cannot be left to voluntary measures. Enforcement is typically lacking of such ethics codes. Apart from that, many self-regulatory AI principles are rather vague and fail to give detailed guidance.[107] Wagner warns for 'ethics washing' in the context of AI. He cautions that firms may see ethics 'as the "easy" or "soft" option'.[108] Indeed, self-regulation should not distract legislators from the possible necessity of new laws. The next section discusses whether and how legal protection against algorithmic discrimination could be improved.

## 5. Improving regulation

There may be a need for additional regulation to protect people against algorithmic discrimination. Section 5.1 below explores how existing non-discrimination norms could be enforced more effectively in the area of algorithmic decision-making. Section 5.2 discusses whether non-discrimination norms themselves should be amended because of algorithmic decision-making. The suggestions below serve as starting points for a discussion, rather than as definitive policy advice.

## 5.1. Enforcement

The overarching norms are reasonably clear regarding algorithmic discrimination. Our societies do not, and should not, accept discrimination on the basis of protected characteristics such as ethnic origin. How could enforcement of non-discrimination norms be improved in the area of algorithmic decision-making?

As discussed, one of the problems with algorithmic systems is their 'black box' character.[109] This opaqueness can be seen as a problem in itself – but the opaqueness also hinders discovering discrimination.

Regulation could help to make algorithmic decision-making more transparent. Regulation could, for example, require that algorithmic systems used in the public sector are developed in such a way that they enable auditing and explainability. Such requirements could also be considered for the private sector.[110] Hence, there is a precedent for transparency requirements in the private sector. In the EU, a legal requirement for interpretability exists for certain systems for algorithmic trading used by investment firms.[111]

Sometimes, public sector bodies could release the underlying code (software) of algorithmic systems. In some circumstances, examining the code can provide information about how a system works. As Rieke, Bogen, and Robinson note, 'code audits are most likely to be useful when there is a clearly defined question about how a software program operates in regulated space, and particular standards against which to measure a system's behavior or performance'.[112] Freedom of information laws could be adapted, so that the code in algorithmic systems is subject to such laws. Such an amendment would enable journalists, researchers, and others to obtain and examine such code.

In many circumstances, examining the code of an algorithmic system does not provide much useful information, as the system can only be assessed when it is used in practice. 'For even moderately complex programs', observe Rieke, Bogen, and Robinson, 'it may be necessary to see a program run "in the wild", with real users and data to truly understand its effects'.[113]

Algorithmic systems are often protected by trade secrets, intellectual property rights, or a firm's terms and conditions. Such protection makes it harder for regulators, journalists, and researchers to investigate such systems. Perhaps the law should be adapted to improve research exceptions. And perhaps the law should require private organisations to disclose certain information to researchers upon request.[114] Such regulation must strike a delicate balance between public interest in transparency and commercial, privacy, and other interests in opaqueness.[115]

The law could require that the public sector only uses algorithmic systems that enable oversight and auditing and that were properly assessed for risks.[116] A similar requirement could be considered for the private sector, when algorithmic systems are used for certain types of decisions, for example about eligibility for insurance, credit, or a job.[117] Considerable expertise is needed for auditing of algorithmic systems; more research and debate are needed on who should conduct such audits.

Equality Bodies and Data Protection Authorities should have sufficient investigation and enforcement powers, and should receive adequate funding, for instance to hire technical expertise.

In conclusion, new regulation should aim for better enforcement of current non-discrimination norms in the area of algorithmic decision-making. However, algorithmic decision-making also enables new types of discrimination and differentiation that could evade current non-discrimination and other laws. We turn to that topic now.

## 5.2. Regulating new types of discrimination and differentiation

Many non-discrimination statutes only apply to certain protected classes (characteristics), such as ethnicity, gender, or sexual orientation.[118] However, as scholars such as Custers,[119] Mantelero,[120] and Wachter[121] note, if an algorithmic system differentiates on the basis of newly invented classes, it could remain outside the scope of non-discrimination law.

To give a simplified example: suppose an algorithmic system finds a correlation between (i) using a certain web browser, and (ii) a higher willingness to pay. An online store could charge higher prices to people using that browser.[122] Such practices would remain outside the scope of non-discrimination law, as a browser type is not a protected characteristic. (For this hypothetical we assume that the browser type is not a proxy for a protected characteristic.)

But algorithmic decisions can be unfair or have other drawbacks, even if the decisions remain outside the scope of non-discrimination law. For example, insurance companies could use algorithmic systems to set premiums for individual consumers, or to deny them insurance. To some extent, risk differentiation is necessary, and an accepted practice, for insurance. And to some extent it is fair when high-risk customers pay higher premiums. However, insurance could become unaffordable for some consumers if insurance companies engage in too much risk classification. Hence, algorithmic classification could threaten the risk-pooling function of insurance.

Furthermore, algorithmic decision-making could reinforce social inequality.[123] Somebody who lives in a poor neighbourhood with many burglaries might pay more for house insurance, because the risk of a burglary is higher. If neighbourhoods where many poor people live have higher risks, then poor people pay, on average, more.[124] However, someone's financial status is not a protected characteristic, so non-discrimination law does not regulate such a practice (assuming that the practice does not lead to indirect discrimination based on a protected characteristic).[125]

In addition, non-discrimination law has little to say about algorithmic predictions that are incorrect (false positives and false negatives). A problem with algorithmic decisions is that they are often incorrect for a particular individual. Algorithmic decision-making often entails applying a predictive model to individuals. A simplified example of a predictive model is: '80% of the people living in postal code 10017 pay their bills late'. Suppose that a firm uses this predictive model and denies loans to all people in postal code 10017. The firm would also deny loans to the 20% who pay their bills on time.[126] Sometimes, algorithmic systems make more errors for minority groups than for the majority. Hence, errors could disproportionately harm certain groups.[127] In conclusion, algorithmic decision-making that evades non-discrimination law can still be unfair.

## 5.3. What should policymakers do?

Additional regulation should be considered to defend human rights and fairness in the area of algorithmic decision-making. It would not be useful, however, to regulate algorithmic decision-making in general; the use of algorithms is too varied for one set of rules.

Algorithmic decision-making is used in many different sectors, for many different purposes – in most cases, algorithmic decision-making does not threaten human rights.[128] An algorithmic system of a chess computer brings different risks, if any, than an algorithmic system for credit rating or predictive policing.

Even for algorithmic systems that make decisions about humans, the risks are different in different sectors, and different rules should apply. We cannot assess the fairness of algorithmic decisions in the abstract. Depending on the sector, or application area, different arguments have different weights, and different normative and legal principles apply. For example, in the field of criminal law, the right to a fair trial and the presumption of innocence are important. In consumer transactions, freedom of contract is an important principle. Therefore, new rules should focus on specific sectors.

To assess whether new rules are needed, the following approach could be followed. For a particular sector, several questions should be answered.

(i) Which rules apply in this sector, and what are their rationales? A rule may, for instance, aim to protect a human right. Or a rule could express a legal principle, such as equality, contractual freedom, or the right to a fair trial. Economic rationales also differ from sector to sector. For example, risk pooling is important for insurance, while it is not relevant in most other sectors. Hence, for each sector the rationales for rules differ.

(ii) What are the risks of algorithmic decision-making in this sector? For example, in some sectors, incorrect predictions could lead to serious problems, while incorrect predictions are less problematic in other sectors. False positives are a serious problem in the context of criminal law. A false positive could lead to people being questioned, arrested, or perhaps even detained. We should not accept algorithmic decisions that breach the underlying values of criminal law, such as the right to a fair trial or the presumption of innocence. By contrast: if an incorrect prediction by an algorithmic system leads to a badly targeted ad on the internet, the effect is often less harmful.

(iii) Considering the rationales for the rules in this sector, should the law be improved in the light of algorithmic decision-making? Does algorithmic decision-making threaten the law's underlying principles, or undermine the law's goals? If current law leaves important risks unaddressed, new rules should be considered.

There is a need for more normative and legal research, and more public debate, about algorithmic discrimination. Are new rules needed, or are tweaks to non-discrimination law and data protection law sufficient? How should the law – and technology – protect people against structural and intersectional discrimination?[129] How to assess fairness in diverse sectors? Are there certain types of decisions that should never be taken by algorithms?

## 6. Concluding thoughts

We can benefit tremendously from algorithmic decision-making. However, algorithmic decision-making brings risks too. Computer scientists have shown how algorithmic systems can discriminate, for example when such a system reproduces discrimination that it learnt from discriminatory human decisions. This paper discussed how the law in Europe should react to the problem of algorithmic discrimination.

The effects of algorithmic decisions can be far-reaching. In the public sector, algorithms can be used for predictive policing or sentencing recommendations, and for decisions about, for example, pensions, housing assistance, or unemployment benefits. In the private sector too, algorithmic decisions can seriously affect people, for instance when decisions concern employment, housing, or credit. Even decisions that each have only small effects, could have major effects together. It may not be a big problem if somebody pays 10% extra for a T-shirt because of online price differentiation. Price differentiation could become a problem, however, if certain groups in society structurally pay more for goods and services.

Non-discrimination law and data protection law are the most relevant legal instruments to fight illegal discrimination by algorithmic systems. If effectively enforced, both legal instruments can help to protect people. The paper suggested how enforcement of non-discrimination norms could be improved.

But some types of algorithmic decisions evade current laws, while they can lead to unfair differentiation or discrimination. For instance, many non-discrimination statutes only apply to discrimination on the basis of certain protected grounds, such as ethnic origin. Such statutes do not apply if organisations differentiate on the basis of newly invented classes that do not correlate with protected grounds. Such differentiation could still be unfair, however, for instance when it reinforces social inequality.

We probably need additional regulation to protect fairness and human rights in the area of algorithmic decision-making. However, it is probably not useful to adopt rules for algorithmic decision-making in general. Just like we did not, and could not, adopt one statute to regulate the industrial revolution, we cannot adopt one statute to regulate algorithmic decision-making. To mitigate problems caused by the industrial revolution, we needed different laws for work safety, consumer protection, the environment, etc. In different sectors, the risks are different, and different norms and values are at stake. Therefore, new rules for algorithmic decision-making should be sector-specific.

## ORCID

*Frederik J. Zuiderveen Borgesius*  http://orcid.org/0000-0001-5803-827X

## Notes

## Acknowledgements

The author wishes to thank Bodó Balázs, Noël Bangma, Janneke Gerards, Irena Guidikova, Dick Houtzager, Margot Kaminski, Dariusz Kloza, Gianclaudio Malgieri, Stefan Kulk, Linnet Taylor, Michael Veale, Sandra Wachter, Bendert Zevenbergen, and the anonymous reviewers for their valuable suggestions.

## Disclosure statement

## Notes on contributor

*Prof Frederik J. Zuiderveen Borgesius* is Professor ICT and Law at Radboud University Nijmegen, where he is affiliated with the interdisciplinary research hub on Security, Privacy, and Data Governance: the iHub. His research interests include privacy, discrimination, and other fundamental rights, especially in the context of new technologies. He tweets at: https://twitter.com/FBorgesius.

## Funding

This work was supported by EU Marie Curie individual grant [grant number 748514, PROFILE].